\documentclass[%
 reprint, 
 superscriptaddress,
 amsmath,amssymb,
 aps, 
 pra,
 floatfix,
 reprint
]{revtex4-2}

\usepackage{graphicx}
\usepackage{dcolumn}
\usepackage{bm}
\usepackage{hyperref}

\usepackage{siunitx}
\DeclareSIUnit\corehour{\text{core-hours}}
\DeclareSIUnit\MeVee{\text{MeVee}}
\DeclareSIUnit\keVee{\text{keVee}}
\DeclareSIUnit[quantity-product = ]\percent{\char`\%}

\usepackage[nameinlink,capitalize]{cleveref}
\newcommand{\myqtyrshort}[3]{\qtyrange[exponent-mode=input,range-exponents=individual,print-unity-mantissa=false]{#1}{#2}{#3}}

\AddToHook{cmd/appendix/before}{\crefalias{section}{appendix}}

\usepackage{upgreek}
\usepackage{nicematrix}
\usepackage{derivative} 
\usepackage{mathtools} 
\usepackage{xspace} 
\usepackage{pifont} 
\usepackage{algorithmicx}
\usepackage{algorithm}
\usepackage{algpseudocode}
\usepackage{makecell}
\usepackage{multirow}
\usepackage[version=4]{mhchem}

\algnewcommand{\Input}[1]{\State \textbf{Input:} #1}
\algnewcommand{\LineComment}[1]{\Statex \(\triangleright\) \textit{#1}}

\newcommand{\mymatht}[1]{\mbox{\ensuremath{#1}}}
\newcommand{\mymathtv}[1]{\ensuremath{#1}}

\newcommand{\myOrdernum}[1]{\ensuremath{\mathcal{O}(\num{#1})}}
\newcommand{\myOrderqty}[2]{\ensuremath{\mathcal{O}(\num{#1})\,\unit{#2}}}

\newcommand{\myincrement}{\ensuremath{\Updelta}}
\newcommand{\myvect}[1]{\ensuremath{\boldsymbol{\mathrm{#1}}}}
\newcommand{\myrandomvec}[1]{\ensuremath{\boldsymbol{\mathit{#1}}}}

\newcommand{\mysum}[2]{\sum\limits_{#1}^{#2}}

\newcommand{\myprod}[2]{\prod\limits_{#1}^{#2}}

\newcommand{\myodd}[2][]{\ensuremath{\odif[#1]{#2}}}

\newcommand{\myNat}{\mathbb{N}}
\newcommand{\myReal}{\mathbb{R}}

\newcommand{\myReall}[1]{\mathbb{R}_{#1}}
\newcommand{\myNatu}[1]{\mathbb{N}^{#1}}
\newcommand{\myRealu}[1]{\mathbb{R}^{#1}}

\newcommand{\myRealul}[2]{\mathbb{R}^{#1}_{#2}}

\newcommand{\mygammaf}[1]{\ensuremath{\operatorname{\Upgamma}\!\left(#1\right)}}

\newcommand{\mycode}[1]{\texttt{#1}}

\newcommand{\myonlinecite}[1]{Ref.~\onlinecite{#1}}
\newcommand{\myonlinecitepl}[1]{Refs.~\onlinecite{#1}}

\newcommand{\myPuBe}{\text{PuBe}}
\newcommand{\myCf}{\text{Cf-252}}
\newcommand{\mynuclt}[3]{\ensuremath{\cramped{\prescript{#1}{#2}{\mathrm{#3}}}}}
\newcommand{\myCfiso}{\mynuclt{252}{98}{Cf}}
\newcommand{\myPuiso}{\mynuclt{239}{94}{Pu}}

\newcommand{\myCsiso}{\mynuclt{137}{55}{Cs}}

\newcommand{\mydispersion}{\ensuremath{\alpha_{\scriptscriptstyle\text{NB}}}}

\newcommand{\mydbtwo}{\ensuremath{\mathcal{D}_1}}
\newcommand{\mydbfour}{\ensuremath{\mathcal{D}_2}}
\newcommand{\mydbseven}{\ensuremath{\mathcal{D}_3}}
\newcommand{\mydbone}{\ensuremath{\mathcal{D}_4}}
\newcommand{\mydbthree}{\ensuremath{\mathcal{D}_5}}
\newcommand{\mydbfive}{\ensuremath{\mathcal{D}_\text{b}}}

\begin{document}

\title{Identifying Neutron Sources using Recoil and Time-of-Flight Spectroscopy
}%

\author{David Breitenmoser}
\email{Lead and contact author: dbreiten@umich.edu}
\author{Ricardo Lopez}%
\author{Shaun D. Clarke}%
\author{Sara A. Pozzi}%
\affiliation{%
Department of Nuclear Engineering \& Radiological Sciences, University of Michigan, 2355 Bonisteel Blvd., Ann Arbor, MI 48109-2104, United States of America
}%

\date{\today}


\begin{abstract}

Neutron-source identification is central to nuclear physics and its applications, from planetary science to nuclear security, yet direct source discrimination from measured neutron spectra remains fundamentally elusive. Here, we introduce a Bayesian protocol that directly infers source ensembles from measured neutron spectra by combining full-spectrum template matching with probabilistic evidence evaluation. Applying this protocol to recoil and time-of-flight spectroscopy, we recover single- and two-source configurations with strong statistical significance ($>\!\!4\sigma$) at event counts as low as $\sim\!\!10^{3}$. These results demonstrate that neutron spectral signatures can be leveraged for robust source identification, opening a new observational window for both fundamental research and operationally driven applications.

\end{abstract}

\maketitle
\newpage

\section{\label{sec:Introduction}Introduction}

Neutrons are powerful probes of matter due to their charge neutrality and strong nuclear interactions \cite{Abele2008,Pietropaolo2020}. The energy spectra of neutron fields encode both their production mechanisms and the modulation processes they undergo, making neutron spectroscopy indispensable for applications ranging from determining planetary surface compositions \cite{Mitrofanov2002a,Feldman1998b,Prettyman2012a,Lawrence2013a} to monitoring environmental neutron fields for soil moisture and biomass estimation \cite{Zreda2008,Franz2013,Andreasen2017}. In safety and security contexts, the ability to infer neutron-source characteristics directly from spectral measurements is equally critical, supporting the discrimination of special nuclear material \cite{Runkle2010,vanderEnde2019} and enabling robust source attribution and risk-informed protective actions in nuclear forensics and emergency response \cite{Mayer2013,Perello2023}.

Despite its central role across scientific and applied domains, inferring neutron sources and the modulation processes they undergo directly from measured energy spectra---referred to here as identification---remains fundamentally elusive. Most neutron sources exhibit continuous energy emissions with strong spectral similarity \cite{Clarke2024a,Perello2023}. These spectral features, compounded by detector response broadening and the modulating effects of intervening materials, result in ill-conditioned inverse problems that make it difficult to determine the underlying source characteristics and modulation histories. As a result, current identification methodologies are largely restricted to indirect or proxy signatures, such as secondary alpha and gamma-ray emissions \cite{Wallenius2007,Mayer2013,Meng2018,Lepowsky2021,LaMont1998} or correlation metrics derived from integral spectral components \cite{Clarke2024a,Feldman1998b,Prettyman2012a,Lawrence2013a}. While secondary emissions can provide additional context, they may be absent or strongly attenuated depending on source encapsulation and environmental conditions. Qualitative spectral metrics, in turn, offer only limited accuracy and precision, are typically restricted to single-source scenarios, and lack quantitative measures of statistical significance.

In this study, we introduce a Bayesian methodology that directly addresses these challenges by enabling quantitative inference of source ensembles using neutron spectroscopy. By combining full-spectrum template matching with probabilistic inference, our approach overcomes the limitations of current methods, providing direct, quantitative discrimination between single- and multi-source models, even in the presence of strong spectral similarity and low count statistics. We demonstrate the identification fidelity in a series of dedicated laboratory experiments using varying sets of spontaneous fission and (\mymatht{\upalpha},n) neutron sources, evaluated across two prominent measurement modalities: recoil and time-of-flight (TOF) spectroscopy.

\section{\label{sec:BayesTheory}Evidence based Bayesian source identification}

Motivated by recent successes in addressing similarly ill-conditioned model inference problems in gravitational-wave astronomy \cite{Veitch2015,Ashton2019,Smith2020} and exoplanet research \cite{MacDonald2017,Pinhas2018}, we develop here a probabilistic framework for neutron source identification formulated as a Bayesian model comparison problem \cite{trotta2008a,vonToussaint2011a}. For a neutron source model \mymatht{\mathcal{M}} conditioned on a set of observed neutron spectra \mymathtv{\mathcal{D}=\{\myvect{y}_i\in\myNatu{N}\}^{N_\mathcal{D}}_{i=1}} with measured counts \mymatht{\myvect{y}\in\myNatu{N}}, \mymatht{N\in\myNat} channels, and \mymatht{N_\mathcal{D}\in\myNat} realizations, we quantify the support for \mymatht{\mathcal{M}} through the model evidence, defined as the marginal likelihood \mymatht{p(\mathcal{D}\mid\mathcal{M})} given by

\begin{equation}
\mathcal{Z} = \int_{\Uptheta} \mathcal{L}( \boldsymbol{\uptheta} ; \mathcal{D} , \mathcal{M}) \, p(\boldsymbol{\uptheta}\mid\mathcal{M})\myodd{\boldsymbol{\uptheta}}
\label{eq:evidence}
\end{equation}

\noindent where \mymatht{\mathcal{L}(\boldsymbol{\uptheta} ; \mathcal{D},\mathcal{M})} and \mymatht{p(\boldsymbol{\uptheta}\mid\mathcal{M})} denote the likelihood and parameter prior defined on a \mymatht{M}-dimensional parameter space \mymatht{\Uptheta=\{\boldsymbol{\uptheta}\in\myRealu{M}\}} with \mymatht{\myNat} and \mymatht{\myReal} representing the sets of natural and real numbers, respectively.

The source model \mymatht{\mathcal{M}} represents the forward mapping from any admissible parameter vector \mymatht{\myvect{\uptheta}} to the expected spectral response of a given spectrometric system \mymathtv{\mathcal{M}(\myvect{\uptheta}) : \Uptheta \subseteq \mathbb{R}^M \mapsto \myNatu{N}}. This mapping is parameterized as a linear superposition of \mymatht{S\in\myNat} spectral templates \mymathtv{\mathcal{M}(\myvect{\uptheta}) = \sum_{s=1}^{S} \xi_s \, \myvect{\uppsi}_s}, where each template \mymathtv{\myvect{\uppsi}_s \in \myRealul{N}{+}} represents the normalized full-spectrum response of the spectrometric system to the \mymatht{s}-th neutron source, and \mymatht{\xi_s \in \mathbb{R}_{+}} denotes the corresponding non-negative neutron emission rate. The template set \mymatht{\mathcal{T} = \{\myvect{\uppsi}_s\}^S_{s=1}} thus defines the discrete ensemble of candidate neutron sources considered within a given model \mymatht{\mathcal{M}}. 

To determine which model is most consistent with the data \mymatht{\mathcal{D}}, we perform Bayesian model comparison across a finite collection of competing models \mymatht{\{\mathcal{M}_k\}_{k=1}^{K}} with associated model priors \mymatht{p(\mathcal{M}_k)}. The relative support for one model over another is quantified by the posterior odds ratio

\begin{equation}
\mathcal{O}_{ij} = \frac{p(\mathcal{M}_i \mid \mathcal{D})}{p(\mathcal{M}_j \mid \mathcal{D})} = \frac{\mathcal{Z}_i}{\mathcal{Z}_j} \, \frac{p(\mathcal{M}_i)}{p(\mathcal{M}_j)} ,
\label{eq:oddsratio}
\end{equation}

\noindent which incorporates both the Bayesian evidence \mymatht{\mathcal{Z}_k} and the model prior \mymatht{p(\mathcal{M}_k)} for model \mymatht{\mathcal{M}_k} \cite{Nelson2020}. In the special case of noncommittal model priors \mymathtv{p(\mathcal{M}_k) = 1/N \;  \forall k}, the posterior odds ratio reduces to the ratio of evidence values, \mymatht{\mathcal{Z}_i / \mathcal{Z}_j}, which is commonly referred to as the Bayes factor \cite{trotta2008a,vonToussaint2011a}. 

While \cref{eq:evidence} is analytically intractable for general likelihood and prior distributions \cite{vonToussaint2011a}, advances in Bayesian computation over the past decades have made accurate numerical approximations feasible. In this work, we employ nested sampling to evaluate \cref{eq:evidence} for a predefined set of source models \mymatht{\{\mathcal{M}_k\}_{k=1}^{K}}, likelihood functions \mymatht{\mathcal{L}( \boldsymbol{\uptheta} ; \mathcal{D},\mathcal{M})}, and corresponding parameter priors \mymatht{p(\boldsymbol{\uptheta}\mid\mathcal{M})} \cite{Ashton2022a}. Here, we summarize the likelihood and prior definitions that constitute the Bayesian model, with detailed computational procedures for sampling, convergence diagnostics, and uncertainty quantification provided in \cref{sec:appbayes}.

Following analogous treatments in particle physics and cosmology \cite{Praszalowicz2011,Tezlaf2023,Perez2021,Fry2013,Hurtado-Gil2017,Hameeda2021}, we formulate the likelihood \mymatht{\mathcal{L}(\boldsymbol{\uptheta} ; \mathcal{D},\mathcal{M})} within a probabilistic negative binomial model \mymatht{\myrandomvec{Y}\mid\myvect{\uptheta}\sim\mathcal{NB}(\myvect{y}\mid\myvect{\uptheta})}, where \mymatht{\myrandomvec{Y}} denotes the random vector of observed counts. The associated likelihood was formulated using a parameterization established in prior studies \cite{Hunnefeld2022a,Salinas2020a,Lloyd-Smith2007a} as

\begingroup
\allowdisplaybreaks
\begin{multline}
    \mathcal{L}\left(\myvect{\uptheta};\mathcal{D},\mathcal{M}\right) = \myprod{k=1}{N_{\mathcal{D}}}\myprod{j=1}{N} \frac{\mygammaf{y_{k,j}+\frac{1}{{\mydispersion}}}}{\mygammaf{\frac{1}{{\mydispersion}}}\mygammaf{y_{k,j}+1}}\cdot\\%
    \left(\frac{1}{1+{\mydispersion}\mathcal{M}(\myvect{\uptheta})}\right)^{\frac{1}{{\mydispersion}}}
    \left(\frac{1}{1+[{\mydispersion}\mathcal{M}(\myvect{\uptheta})]^{-1}}\right)^{\mathrlap{y_{k,j}}}\hspace{2mm}\label{eq:likeli}
\end{multline}
\endgroup

\noindent with \mymatht{\mygammaf{\cdot}} denoting the gamma function and \mymatht{y_{k,j}\in\myNat} representing the counts observed in channel \mymatht{j} of the spectrum \mymatht{k}. The dispersion parameter \mymatht{\mydispersion \in \myReall{+}} in \cref{eq:likeli} controls the overdispersion beyond the intrinsic Poisson statistics as \mymatht{\operatorname{Var}(\myrandomvec{Y})=\langle\myrandomvec{Y}\rangle+\mydispersion\langle\myrandomvec{Y}\rangle^2} with \mymatht{\mydispersion\in\myReall{+}}. To prevent biased inference, we incorporate the dispersion parameter \mymatht{\mydispersion} directly into the inference protocol, estimating it concurrently with the neutron emission rates \mymatht{\myvect{\xi}\in \myRealul{S}{+}}. As a result, the inverse problem associated with a given model \mymatht{\mathcal{M}} becomes \mymatht{M=S+1} dimensional, with the full parameter vector defined as \mymatht{\myvect{\uptheta} \coloneqq (\myvect{\xi}, \mydispersion)}.

The likelihood specification in \cref{eq:likeli} forms the first component of the Bayesian model definition. Completing the model requires assigning suitable priors to the model parameters \mymatht{\myvect{\uptheta}}. To avoid overly restrictive assumptions on these parameter priors, we adopted weakly informative, statistically independent marginal priors $p\left(\myvect{\uptheta}\right) \coloneqq \prod_{i=1}^{M}p\left( \theta_{i}\right)$ for all model parameters \mymatht{\theta}, with the marginal priors \mymatht{p\left( \theta\right)} defined based on the principle of maximum entropy \cite{Jaynes1957b}. Specifically, we chose a truncated normal distribution for the neutron emission rate \mymathtv{\xi\sim\mathcal{N}(\mu=\qty{1d8}{\per\s},\sigma=\qty{1d8}{\per\s})\in [0,\infty)}, invariant with respect to the source, and an exponential distribution for the dispersion parameter \mymathtv{\mydispersion\sim\mathcal{E}(\lambda=1)\in [0,\infty)}.

\section{\label{sec:Measurement}Experiments}

We demonstrate the fidelity of the proposed Bayesian identification methodology by retrieving spontaneous fission and (\mymatht{\upalpha},n) source combinations from observed neutron fields probed by neutron spectroscopy under laboratory conditions. 

\begin{figure}[tb]%
\centerline{
\includegraphics[width=1\linewidth]{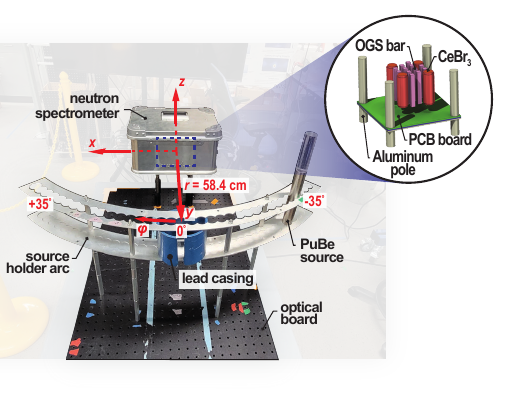}}
\caption[Experimental setup]{Measurement setup for experiment {\mydbfour} (single {\myPuBe} source at azimuth \mymatht{\varphi=\qty{-35}{\degree}} relative to the spectrometer, see \cref{tab:measurement}).}\label{fig:setup}
\end{figure}

The neutron fields in all experiments were measured using an array of twelve organic-glass scintillator bars, each coupled to a dual-ended silicon-photomultiplier readout \cite{Lopez2022}, enabling simultaneous recoil and TOF spectroscopy with a single detector system. Three independent experiments were performed covering the complete power set of a {\myCfiso} and a {\myPuiso}-Be source, hereafter referred to as {\myCf} and {\myPuBe} with respective neutron emission rates of \mymatht{\qty{1.74(9)d6}{\per\s}} and \mymatht{\qty{1.70(4)d6}{\per\s}}. The sources were fixed at a distance of \qty{58.4}{\cm} from the center of the spectrometer's active volume, centered on its horizontal mid-plane. Reproducible placement of sources and the spectrometer was ensured using an optical breadboard and a custom source holder arc (see \cref{fig:setup}).

A dedicated background run was performed under identical detector and acquisition conditions to quantify the intrinsic neutron background in both recoil and TOF modes. The background contribution was found to be negligible in all cases, with signal‑to‑background ratios exceeding \num{1d4} in recoil spectroscopy and \num{1d3} in TOF spectroscopy (see Supplemental Material \cite{zotero-item-8041}).

\nocite{Foreman-Mackey2016}

Templates \mymatht{\myvect{\uppsi}} for the measured sources were inferred from two additional single-source measurements using a data-driven approach based on generalized additive machine-learning models \cite{Serven2025}. To test the robustness and generalization capability of the derived template bank, we deliberately varied the azimuthal angle \mymatht{\varphi} and introduced an additional \qty{6.8}{\mm} lead sleeve during these measurements. A summary of all source–detector configurations and corresponding live times is provided in \cref{tab:measurement}. Further details on the spectrometer, data reduction pipeline, and template generation protocol can be found in \cref{sec:appspec,sec:apppost,sec:apptemplate}.

\begin{table}[tb]
\caption[Performed neutron spectroscopy experiments]{\label{tab:measurement}In this table, we summarize key measurement parameters of the performed neutron spectroscopy experiments.}
\begin{ruledtabular}
\begin{tabular}{c c S[detect-weight,exponent-mode=scientific] c S[detect-weight,exponent-mode=scientific] c}
ID & Source & \multicolumn{1}{c}{\mymatht{\xi} (\unit{\per\s})\footnotemark[1]} & \multicolumn{1}{c}{\mymatht{\varphi} (\unit{\degree})\footnotemark[2]} & \multicolumn{1}{c}{\mymatht{t_\text{live}} (\unit{\s})\footnotemark[3]} & Casing\footnotemark[4] \\
\hline
{\mydbtwo}                  & {\myCf}                & \num{1.74(9)d6}      & $+35$   & \num{5.63d4}  & \ding{55}\\
{\mydbfour}                 & {\myPuBe}              & \num{1.70(4)d6}      & $-35$   & \num{1.80d4}  & \ding{55}\\
{\mydbseven}                & {\myCf}                & \num{1.74(9)d6}      & $+35$   & \num{2.03d4}  & \ding{55}\\
                            & {\myPuBe}              & \num{1.70(4)d6}      & $-35$   &               & \ding{55}\\      
{\mydbone}                  & {\myCf}                & \num{1.74(9)d6}      & $0$     & \num{5.63d4}  & \ding{51}\\
{\mydbthree}                & {\myPuBe}              & \num{1.70(4)d6}      & $0$     & \num{1.80d4}  & \ding{51}\\
{\mydbfive}                 & \textit{background}    &                      &         & \num{5.32d4}  &          \\
\end{tabular}
\end{ruledtabular}
\footnotetext[1]{Neutron emission rate.}
\footnotetext[2]{Azimuth source position with respect to the spectrometer (see \cref{fig:setup}).}
\footnotetext[3]{Measurement live time (rounded to three significant figures).}
\footnotetext[4]{Flag indicating use of a lead sleeve (\qty{6.8}{\mm} thickness) around the deployed neutron source(s) (see \cref{fig:setup}).}
\end{table}

\section{\label{sec:MainResults}Identification Fidelity}

\begin{figure}
\includegraphics[scale=1]{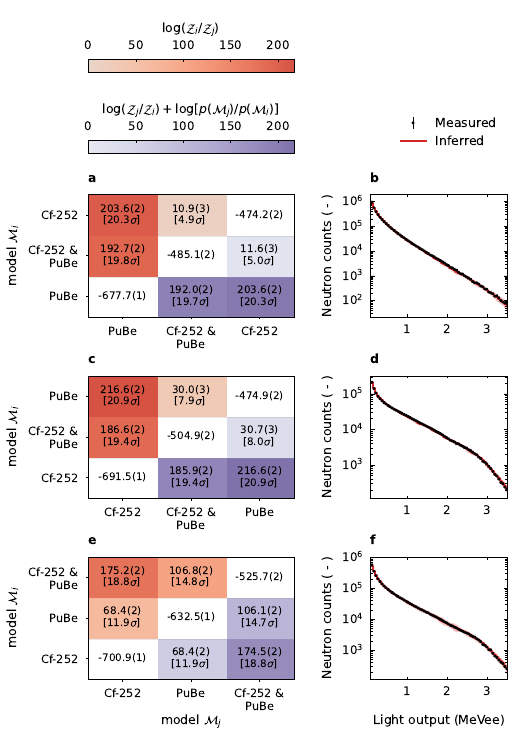}
\caption{\label{fig:ResultRecoil}Bayesian evidence results for three experiments using recoil spectroscopy. (a--b) Single-source Cf-252 experiment ({\mydbtwo}). (c--d) Single-source PuBe experiment ({\mydbfour}). (e--f) Two-source Cf-252 \& PuBe experiment ({\mydbseven}). Panels (a,c,e) show log evidence values \mymatht{\log\mathcal{Z}_{i=j}} for each model \mymatht{\mathcal{M}_{i=j}} (anti-diagonal entries), log Bayes factors \mymathtv{\Delta\log\mathcal{Z}_{ij}=\log(\mathcal{Z}_i/\mathcal{Z}_j)} (above anti-diagonal entries), and log posterior odds ratios \mymathtv{\log\mathcal{O}_{ji}=\log(\mathcal{Z}_{j}/\mathcal{Z}_{i})+\log [p(\mathcal{M}_j)/ p(\mathcal{M}_i)]} (below anti-diagonal entries) with \mymathtv{p(\mathcal{M}) \propto 4^{-\operatorname{dim}(\mathcal{M})}}. Uncertainties are indicated using least-significant-figure notation, with lower statistical-significance bounds in square brackets \cite{trotta2008a}. Panels (b,d,f) show the measured energy spectra (coverage factor \mymatht{k=1}) alongside maximum-a-posteriori predictions and \qty{99}{\percent} central posterior predictive intervals (shaded area) for the retrieved (true) source set.}
\end{figure}

\begin{figure}
\includegraphics[scale=1]{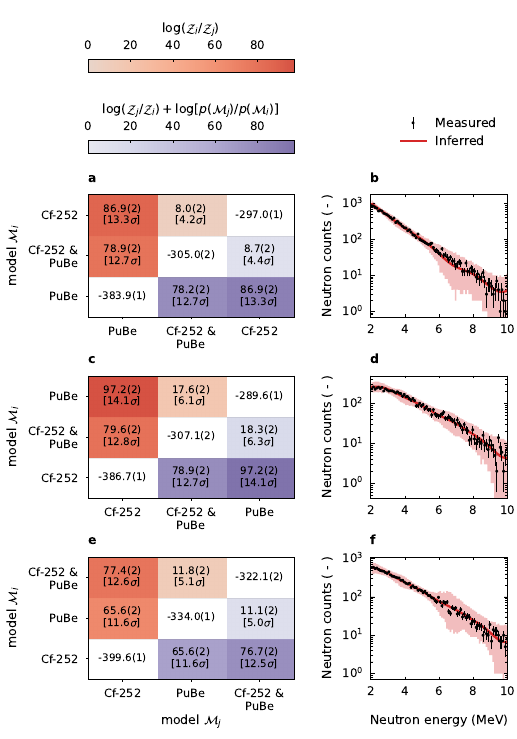}
\caption{\label{fig:ResultTOF}Same as \cref{fig:ResultRecoil} but using TOF instead of recoil spectroscopy.}
\end{figure}

We report the Bayesian evidence results for the complete set of competing models, defined as the power set of single-source templates, \mymatht{\mathcal{P}(\mathcal{S})=\{\mathcal{M}_k\}_{k=1}^{K}}, where \mymatht{\mathcal{S}} denotes the set of individual single-source templates. Results are shown for the three identification experiments in \cref{fig:ResultRecoil,fig:ResultTOF} using recoil and TOF spectroscopy. To illustrate the robustness of our results across retrieval assumptions, we explore the model-selection fidelity using both log Bayes factors \mymatht{\myincrement\log\mathcal{Z}}, representing noncommittal model prior assumptions, as well as log posterior odds ratios \mymatht{\log\mathcal{O}}, adopting weakly informative model priors \mymatht{p\left(\mathcal{M}\right) \propto 4^{-\operatorname{dim}\left(\mathcal{M}\right)}} favoring lower-dimensional models \cite{Cloutier2019,Nelson2020}. For both spectroscopy modalities, the true source set is recovered with decisive statistical confidence on Jeffreys’ scale (\mymatht{\myincrement\log\mathcal{Z}>\num{8.0}}, \mymatht{\log\mathcal{O}>\num{8.7}}) over all competing models \cite{Jeffreys1948}, corresponding to a statistical significance \mymatht{>\!\!4\upsigma} \cite{trotta2008a}. 

Comparing noncommittal and weakly informative model priors, we observe statistically significant differences in the inferred evidence for the true source configuration, which nevertheless do not alter the model-selection outcome. For example, in the recoil spectroscopy modality of the {\myCf} \& {\myPuBe} experiment (\cref{fig:ResultRecoil}e), the log posterior odds ratio (weakly informative prior) is \num{106.1(2)}, while the corresponding log Bayes factor (noncommittal prior) is \num{106.8(2)} relative to the best competing model ({\myPuBe}), illustrating that the model prior choice has a negligible effect on the model selection fidelity, and confirming that the outcome is dictated by the observed data rather than by the specific form of the model priors for the given measurement setup. 

Here, all identification results were obtained with a fixed measurement time \mymatht{\myOrderqty{1d4}{\s}} per experiment, with recoil and TOF spectroscopy acquisitions performed in parallel. Because TOF spectroscopy relies on coincidence events, its event rate is \mymatht{\myOrdernum{1d2}} lower than that of recoil spectroscopy, leading to consistently higher log Bayes factors and log posterior odds ratios for the recoil modality. To generalize beyond system-specific event rates, we quantify in the next section the scaling of model-selection fidelity with the total number of detected events for both spectroscopy modalities.

\section{\label{sec:timeresult}Event scaling}

\begin{figure}
\includegraphics[scale=1]{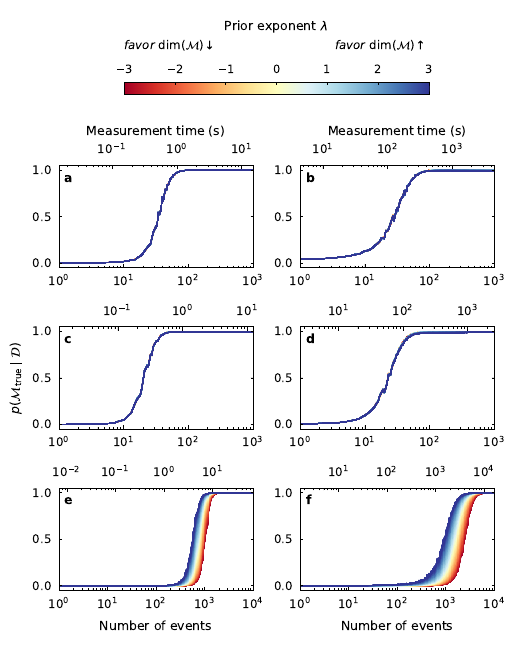}
\caption{\label{fig:ResultPMP}Posterior model probability \mymatht{p(\mathcal{M}_\mathrm{true}\mid\mathcal{D})} of retrieving the true source model \mymatht{\mathcal{M}_\mathrm{true}} as a function of the total number of detected events, for varying prior exponents \mymatht{\lambda} in the model prior \mymathtv{p(\mathcal{M}) \propto 2^{\lambda\operatorname{dim}(\mathcal{M})}} with \mymatht{\lambda\in[-3,+3]}. Panels (a,b) show the single-source Cf-252 experiment ({\mydbtwo}), (c,d) the single-source PuBe experiment ({\mydbfour}), and (e,f) the two-source Cf-252 \& PuBe experiment ({\mydbseven}). Panels (a,c,e) display posteriors conditioned on recoil spectroscopy data, while (b,d,f) show posteriors conditioned on TOF spectroscopy data. In addition to the number of detected events, the corresponding measurement times for each spectroscopy modality are indicated.}
\end{figure}

To assess the scaling of model-selection fidelity with the total number of detected events, we compute the posterior probability \mymatht{p(\mathcal{M}_\mathrm{true}\mid\mathcal{D})} of retrieving the true source model \mymatht{\mathcal{M}_\mathrm{true}} conditioned on the experimental data \mymatht{\mathcal{D}} for varying event counts (see \cref{sec:appbayes}). Additionally, to explore the impact of the applied prior assumptions on \mymatht{p(\mathcal{M}_\mathrm{true}\mid\mathcal{D})}, we relax the model prior used previously to \mymathtv{p(\mathcal{M}) \propto 2^{\lambda\operatorname{dim}(\mathcal{M})}} with \mymatht{\lambda\in[-3,+3]} and \mymatht{\lambda=0} representing a noncommittal model prior. In \cref{fig:ResultPMP}, the results of these computations for recoil and TOF spectroscopy are presented. As expected, the experiments differ in how quickly they decisively favor the true model, with the single-source cases achieving \mymatht{p(\mathcal{M}_\mathrm{true}\mid\mathcal{D})\approx 1} after \num{\sim1d2} events, whereas the two-source cases require \num{\sim1d3} events. Owing to the natural regularization inherent in the Bayesian evidence, we observe that increasing the model prior exponent \mymatht{\lambda} consistently leads to faster convergence by favoring higher-dimensional models, while decreasing it slows convergence by favoring lower-dimensional models, consistent with the dimensionality weighting in the prior. Interestingly, recoil spectroscopy converges consistently faster than TOF spectroscopy, not only in temporal terms, as discussed above, but also with respect to the absolute event count. This difference in scaling motivates an information-theoretic analysis of the two modalities to determine whether one yields an inherently higher information gain than the other on an absolute event scale.

\begin{figure}
\includegraphics[scale=1]{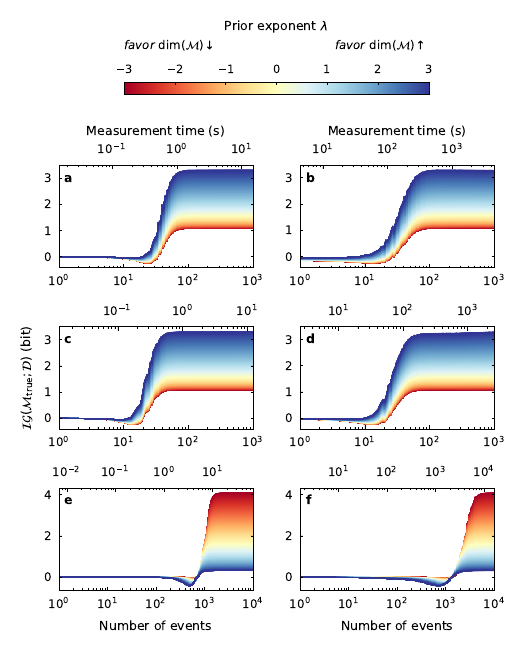}
\caption{\label{fig:ResultInfo}Information gain \mymatht{\mathcal{IG}\left(\mathcal{M}_\mathrm{true};\mathcal{D}\right)} about the true source model \mymatht{\mathcal{M}_\mathrm{true}} from the dataset \mymatht{\mathcal{D}} as a function of the total number of detected events, for varying prior exponents \mymatht{\lambda} in the model prior \mymathtv{p(\mathcal{M}) \propto 2^{\lambda\operatorname{dim}(\mathcal{M})}} with \mymatht{\lambda\in[-3,+3]}. Panels (a,b) show the single-source Cf-252 experiment ({\mydbtwo}), (c,d) the single-source PuBe experiment ({\mydbfour}), and (e,f) the two-source Cf-252 \& PuBe experiment ({\mydbseven}). Panels (a,c,e) display \mymatht{\mathcal{IG}} for recoil spectroscopy data, while (b,d,f) show \mymatht{\mathcal{IG}} for TOF spectroscopy data. In addition to the number of detected events, the corresponding measurement times for each spectroscopy modality are indicated.}
\end{figure}

Following an information-theoretic perspective \cite{Lindley1956}, we aim to quantify the information gained \mymatht{\mathcal{IG}\left(\mathcal{M}_\mathrm{true};\mathcal{D}\right)} about the true source model \mymatht{\mathcal{M}_\mathrm{true}} from our experimental data \mymatht{\mathcal{D}} for varying event counts relative to our prior beliefs \mymatht{p(\mathcal{M}_\mathrm{true})} over all candidate models. Building on earlier information theory work in Bayesian parameter inference \cite{Lindley1956,Luttrell1985,MacKay1992}, we define this information gain as the relative entropy \cite{Kullback1951}, which measures the information (in bits) gained about the true source model \mymatht{\mathcal{M}_\mathrm{true}} from the dataset \mymatht{\mathcal{D}} (see \cref{sec:appbayes}). In the limit where the posterior collapses on the true model \mymatht{p(\mathcal{M}_\mathrm{true}\mid\mathcal{D})\to 1}, \mymatht{\mathcal{IG}(\mathcal{M}_\mathrm{true};\mathcal{D})} approaches \mymatht{-\log_2 p(\mathcal{M}_\mathrm{true})}. Thus, the prior places an absolute upper bound on the achievable information. In \cref{fig:ResultInfo}, we show \mymatht{\mathcal{IG}(\mathcal{M}_\mathrm{true};\mathcal{D})} for all three experiments as a function of event count. Across all tested prior weights and source configurations, recoil spectroscopy exhibits a consistently steeper rise in information gain than TOF spectroscopy, demonstrating that---beyond the expected advantage from its higher event rate---recoil spectroscopy also achieves a higher information-acquisition rate on an absolute event scale. Based on both parameter and predictive posteriors across all experiments, we attribute the lower information gain in TOF spectroscopy to its substantially higher overdispersion relative to recoil spectroscopy (factor \mymatht{\num{\sim1d1}} difference in the dispersion parameter \mymatht{\mydispersion}), resulting in an inherently higher statistical dispersion per event. Full posterior results for all experiments are provided in the Supplemental Material \cite{zotero-item-8041}.

\begin{figure}
\includegraphics[scale=1]{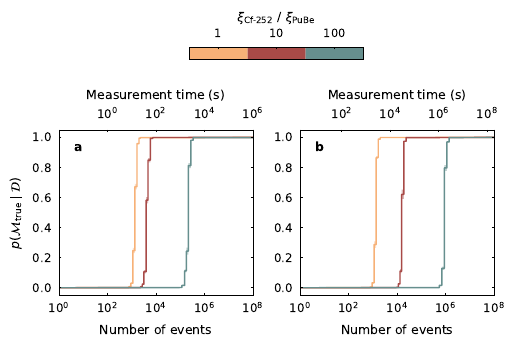}
\caption[Scaling with source emission ratio]{\label{fig:ResultRatio}Posterior model probability \mymatht{p(\mathcal{M}_\mathrm{true}\mid\mathcal{D})} of retrieving the true source model \mymatht{\mathcal{M}_\mathrm{true}} ({\myCf} \& {\myPuBe}) as a function of the total number of detected events, for varying neutron emission ratios \mymatht{\xi_{\myCf}/\xi_{\myPuBe}} with non-commital model priors \mymathtv{p(\mathcal{M}) =\text{const}} (corresponding to \mymatht{\lambda=0}). Panel (a) displays posteriors conditioned on recoil spectroscopy data, while panel (b) shows posteriors conditioned on TOF spectroscopy data. For reference, we indicate, in addition to the number of detected events, measurement times for each spectroscopy modality assuming a neutron emission rate \mymatht{\xi_{\myPuBe}=\qty{1.70(4)d6}{\per\s}} for {\myPuBe} reflective of the one used in our experiments (see \cref{sec:Measurement}). Uncertainties are indicated as 3-sigma shaded areas.}
\end{figure}

We conclude the event-scaling analysis by examining how the posterior probability \mymatht{p(\mathcal{M}_\mathrm{true}\mid\mathcal{D})} of retrieving the true source model \mymatht{\mathcal{M}_\mathrm{true}} scales with the total number of detected events when the neutron emission ratio in our experiments \mymatht{\xi_{\myCf}/\xi_{\myPuBe}\approx1} is relaxed (see \cref{sec:Measurement}). To this end, we generated two-source  datasets by combining resampled {\myCf} \& {\myPuBe} single-source spectroscopic data with varying emission ratios using a resampling protocol described in \cref{sec:appbayes}, in direct analogy to the analysis of \cref{fig:ResultPMP}. The resulting posterior probabilities are shown in \cref{fig:ResultRatio}. As expected, greater imbalance between the neutron emission rates requires more events for the posterior to decisively favor the true model. Nevertheless, even under significant imbalances  \mymatht{\xi_{\myCf}/\xi_{\myPuBe}=\myOrdernum{1d2}}, the correct source set is recovered with decisive statistical confidence on Jeffreys’ scale (\mymatht{\myincrement\log\mathcal{Z}>\num{13}}, \mymatht{>\!\!5\sigma}) after \myOrdernum{1d6} detected events for both spectroscopy modalities, demonstrating the robustness of the Bayesian identification procedure even under strongly non‑uniform source contributions. Consistent with the findings above, recoil spectroscopy reaches decisive posterior probabilities at systematically lower event counts than TOF spectroscopy, indicating that its advantage persists even when the constraint of comparable neutron emission rates is relaxed.

\section{\label{sec:Conclusion}Conclusion}

In this study, we have demonstrated the first quantitative identification of neutron sources from measured neutron spectra by combining full-spectrum template matching with probabilistic evidence evaluation. Our framework achieves high-confidence discrimination (\mymatht{>\!\!4\upsigma}) in single- and two-source configurations with comparable neutron emission rates at event counts as low as \mymatht{\num{\sim1d3}}. Furthermore, we find that this identification remains robust to source imbalances: even at neutron emission ratios of \myOrdernum{1d2}, the true source model is retrieved with decisive statistical confidence within \myOrdernum{1d6} events. Comparing recoil and TOF modalities, we find that recoil spectroscopy delivers consistently higher information gain across all source configurations, both on a temporal and on a per-event basis. This work demonstrates that neutron spectral signatures can be leveraged as a direct and quantitative modality for source identification, opening a new, robust observational window for both basic research and field applications.

While our study focused on simplified source sets under controlled laboratory conditions, the framework naturally extends to more complex and realistic source scenarios. To scale these methods efficiently without the need for prohibitively large measurement campaigns, future work will explore the use of high-fidelity neutron transport simulations \cite{Pozzi2003,Allison2016,Ahdida2022a} to generate comprehensive, physics-informed template banks. By incorporating spectral variability arising from different geometries, shielding materials, or environmental conditions, such physics-informed modeling will enable the joint quantitative inference of both the source ensembles and the modulation processes the neutrons undergo. This capability will extend our retrieval methods to a broad variety of applications, from constraining the surface composition of planetary bodies in the solar system \cite{Mitrofanov2002a,Feldman1998b,Prettyman2012a,Lawrence2013a} to supporting discrimination of special nuclear material \cite{Runkle2010,vanderEnde2019} and robust source attribution in nuclear forensics, radiochemistry, and emergency response \cite{Mayer2013,Perello2023}.

\section{Acknowledgments}
This work was funded in-part by the Consortium for Monitoring, Technology, and Verification under Department of Energy National Nuclear Security Administration Award No. DE-NA0003920.\nocite{Breitenmoser2025h}

\appendix

\section{\label{sec:appbayes}Bayesian Computations}

In this appendix, we describe the numerical evaluation of the Bayesian evidence in \cref{eq:evidence} using nested sampling alongside a thorough discussion of the convergence diagnostics and uncertainty quantification. We also provide details on the Bayesian computations for the event-scale analysis in \cref{sec:timeresult}, including the evaluation of posterior model probabilities and information-gain measures.

\subsection{\label{subsec:appevidence}Bayesian Evidence}

Using the specified joint prior distribution and likelihood function in \cref{sec:BayesTheory}, we evaluated the Bayesian evidence \mymatht{\mathcal{Z}} in \cref{eq:evidence} while simultaneously sampling the posterior distribution \mymatht{p(\myvect{\uptheta}\mid\mathcal{D},\mathcal{M})} using the nested sampling code \mycode{dynesty} (version~\mycode{3.0.0}) \cite{Speagle2020a}. This Bayesian inference procedure was carried out independently for each candidate source model \mymatht{\{\mathcal{M}_k\}_{k=1}^{K}}, yielding a corresponding set of Bayesian evidence values \mymatht{\{\mathcal{Z}_k\}_{k=1}^{K}} from which Bayes factors and posterior odds ratios are readily obtained as detailed in \cref{sec:BayesTheory}. For each nested-sampling run, we used \mymatht{2^{10}} live points and set the evidence tolerance to \mymatht{\Delta \log \mathcal{Z} = 0.1}. The bounding strategy \mycode{single} was adopted, and new samples were drawn using the default uniform-sampling method. All other sampler settings were kept at their default values. Convergence and numerical precision were evaluated using established nested-sampling diagnostics \cite{Higson2019a,Speagle2020a,Ashton2022a}. Across all runs, we obtained an effective sample size \mymatht{\mathrm{ESS} > \num{4d3}} and a sampling efficiency exceeding \qty{25}{\percent}. Following the uncertainty quantification framework introduced in \myonlinecitepl{Skilling2006a,Higson2018a,Higson2019a}, the statistical uncertainties on the computed Bayesian evidence values were estimated directly from the \mycode{dynesty} outputs and are reported in \cref{sec:MainResults}. In addition to the Bayesian evidence values, Bayes factors, and posterior odds ratios reported in \cref{sec:MainResults}, we provide the full posterior results for all nested sampling runs in the Supplemental Material \cite{zotero-item-8041}.

It is important to note that the gamma function terms in \cref{eq:likeli} increase rapidly for moderately large arguments, requiring careful numerical evaluation of the likelihood function. To prevent overflow, we conducted all Bayesian computations using the logarithm of the likelihood function:

\begingroup
\allowdisplaybreaks
\begin{align}
&\log{\mathcal{L}\left(\myvect{\uptheta};\mathcal{D},\mathcal{M}\right)} = \mysum{k=1}{N_{\mathcal{D}}}\mysum{j=1}{N}\log{\mygammaf{y_{k,j}+\frac{1}{{\mydispersion}}}}-\nonumber\\%
&\hspace{15mm}\log{\mygammaf{\frac{1}{{\mydispersion}}}}-
    \log{\mygammaf{y_{k,j}+1}}-\nonumber\\%
&\hspace{25mm}\frac{1}{{\mydispersion}}\log{\left(1+{\mydispersion}\mathcal{M}(\myvect{\uptheta})\right)}-\nonumber\\%
&\hspace{35mm}y_{k,j}\log{\left(1+\frac{1}{{\mydispersion}\mathcal{M}(\myvect{\uptheta})}\right)}.\label{eq:loglikeli}
\end{align}
\endgroup

\noindent The log-gamma terms in \cref{eq:loglikeli} were computed using a numerically robust algorithm \cite{Hare1997,Virtanen2020}, which maintains stability and precision across the full dynamic range of our input arguments.

\subsection{\label{subsec:apppmp}Posterior Model Probability}

In \cref{sec:timeresult}, we evaluated how the posterior model probability \mymatht{p(\mathcal{M}_{\mathrm{true}}\mid\mathcal{D})} and the information gain \mymatht{\mathcal{IG}(\mathcal{M}_{\mathrm{true}};\mathcal{D})} relative to the true source model \mymatht{\mathcal{M}_{\mathrm{true}}} scale with the total number of detected events in the performed experiments. In the following two sections, we detail the computational procedure used to obtain these quantities.

We started our event-scale analysis by generating a collection of resampled datasets 
\mymatht{\{\mathcal{D}^{(b)}=\{\myvect{y}^{(b)}\in \myNatu{N}\}\}_{b=1}^{B}}, where each spectrum 
\mymathtv{\myvect{y}^{(b)}} contained exactly 
\mymatht{N^{(b)}_{\mathrm{event}}} total counts drawn from a predefined reference dataset \mymatht{\mathcal{D}} corresponding to one of the recorded spectra \mymathtv{\myvect{y}\in\myNatu{N}} presented in \cref{sec:MainResults}. Defining the spectral probability vector \mymathtv{\mathbf{p}=(p_1,\ldots,p_N)} with components \mymathtv{p_i = y_i \big/ \sum_{j=1}^{N} y_j}, each spectrum was generated via multinomial sampling \mymathtv{\myvect{y}^{(b)}\allowbreak\sim\allowbreak\operatorname{Mul\-ti\-no\-mi\-al}(\allowbreak N^{(b)}_{\mathrm{event}},\allowbreak\mathbf{p} ) \in \myNatu{N}} with probabilities \mymatht{\mathbf{p}} proportional to the reference spectrum \mymatht{\myvect{y}}. This procedure yielded independent finite-statistics realizations that preserve the spectral shape of \mymatht{\myvect{y}}. For each number of events \mymatht{N^{(b)}_{\mathrm{event}}}, we generated \mymathtv{N_\text{rep}=\num{1d3}} realizations, with \mymatht{N^{(b)}_{\mathrm{event}}} defined on a logarithmically spaced grid with \num{4d2} instances between \mymathtv{N^{(1)}_{\mathrm{event}}=1} and \mymathtv{N^{(B)}_{\mathrm{event}}=N_\text{event}}, with \mymatht{N_\text{event}} being the number of detected events in the reference spectrum \mymatht{\myvect{y}}.

We continued by computing the log Bayesian evidence \mymatht{\log\mathcal{Z}^{(b)}} for each resampled dataset \mymatht{\mathcal{D}^{(b)}} as

\begingroup
\allowdisplaybreaks
\begin{subequations}
\begin{align}  
&\log\mathcal{Z}^{(b)} = \log\int_{\Theta} \mathcal{L}(\myvect{\uptheta};\mathcal{D}^{(b)},\mathcal{M})p(\myvect{\uptheta}\vert\mathcal{M})\odif{\boldsymbol{\uptheta}}\\
&\hspace{1mm}=\log\mathcal{Z}-\log\int_{\Theta}  \underbracket{\frac{\mathcal{L}(\myvect{\uptheta};\mathcal{D}^{(b)},\mathcal{M})}{\mathcal{L}(\myvect{\uptheta};\mathcal{D},\mathcal{M})}}_{\text{likelihood ratio}}
\underbracket{p(\myvect{\uptheta}\vert\mathcal{D},\mathcal{M})}_{\text{posterior}}
\odif{\boldsymbol{\uptheta}}\\
&\hspace{1mm}\approx \log\mathcal{Z} -\log N_\theta +\log\sum_{i=1}^{N_\theta}\exp\left[\phantom{\log\mathcal{L}(\myvect{\uptheta}_i;\mathcal{D}^{(b)},\mathcal{M})-\log\mathcal{L}(\myvect{\uptheta}_i;\mathcal{D},\mathcal{M})}\right.\nonumber\\%
&\hspace{1.5cm}\left.\log\mathcal{L}(\myvect{\uptheta}_i;\mathcal{D}^{(b)},\mathcal{M})-\log\mathcal{L}(\myvect{\uptheta}_i;\mathcal{D},\mathcal{M})\right]\label{eq:logZb}
\end{align}
\end{subequations}
\endgroup

\noindent with \mymatht{\log\mathcal{Z}} denoting the log Bayesian evidence of the reference dataset \mymatht{\mathcal{D}}, while \mymatht{\log \mathcal{L}(\myvect{\uptheta}_i;\mathcal{D},\mathcal{M})} and \mymatht{\log \mathcal{L}(\myvect{\uptheta}_i;\mathcal{D}^{(b)},\mathcal{M})} are the log likelihoods for \mymatht{\mathcal{D}} and \mymatht{\mathcal{D}^{(b)}} evaluated at each posterior sample \mymatht{\{\myvect{\uptheta}_i\}^{N_\theta}_{i=1}} obtained from the nested sampling runs (see \cref{subsec:appevidence}). The importance sampling procedure in \cref{eq:logZb} allowed us to efficiently estimate the log Bayesian evidence for each of the \mymatht{400\times1000} resampled spectra, from which we then computed the mean and standard error using Monte Carlo error propagation across the \mymatht{N_\text{rep}=\num{1d3}} realizations.

Based on the computed set of log Bayesian evidence values \mymatht{\log\mathcal{Z}^{(b)}} for each resampled dataset \mymatht{\mathcal{D}^{(b)}}, we computed the posterior model probability, i.e., the posterior probability \mymatht{p(\mathcal{M}_\mathrm{true}\mid\mathcal{D}^{(b)})} of retrieving the true source model \mymatht{\mathcal{M}_\mathrm{true}} conditioned on \mymatht{\mathcal{D}^{(b)}} and a set of competing models \mymatht{\{\mathcal{M}_k\}_{k=1}^{K}} with associated model priors \mymatht{p(\mathcal{M}_k)}:

\begin{equation}
    p(\mathcal{M}_\text{true} \vert \mathcal{D}^{(b)})=\frac{p(\mathcal{D}^{(b)}\vert\mathcal{M}_\text{true}) p(\mathcal{M}_\text{true})}{\sum^{K}_{k=1}p(\mathcal{D}^{(b)}\vert\mathcal{M}_k) p(\mathcal{M}_k)}\label{eq:pmodel}
\end{equation}

\noindent with \mymatht{p( \mathcal{D}^{(b)}\mid\mathcal{M}_k)=\mathcal{Z}^{(b)}_k} denoting the Bayesian evidence of \mymatht{\mathcal{D}^{(b)}} conditioned on model \mymatht{\mathcal{M}_k}. For numerical stability, \cref{eq:pmodel} was evaluated in logarithmic form using the LogSumExp (LSE) transformation, which allows for robust computation of the logarithm of a sum of exponentials and prevents over- or underflow:  

\begin{multline}
    \log p(\mathcal{M}_\text{true} \vert \mathcal{D}^{(b)})=\log\mathcal{Z}^{(b)}_\text{true} + \log p(\mathcal{M}_\text{true}) -\\
    \log \mysum{k=1}{K}\exp\left[\log\mathcal{Z}^{(b)}_k+\log p(\mathcal{M}_k)\right]\label{eq:logpmodel}
\end{multline}

\noindent where \mymatht{\mathcal{Z}^{(b)}_\text{true}\coloneqq p( \mathcal{D}^{(b)}\mid\mathcal{M}_\text{true})} denotes the Bayesian evidence of \mymatht{\mathcal{D}^{(b)}} conditioned on the true model \mymatht{\mathcal{M}_\text{true}}. As with the Bayesian evidence itself, the statistical uncertainty in the posterior model probability \mymatht{p(\mathcal{M}_\mathrm{true}\mid\mathcal{D}^{(b)})} was quantified via Monte Carlo error propagation over the \mymatht{N_\text{rep}=\num{1d3}} available realizations. 

The computational protocol described above was further generalized to evaluate the source configurations presented in \cref{fig:ResultRatio}. For these evaluations, resampled datasets \mymathtv{\mathcal{D}^{(b)}} were constructed by combining single-source spectral components with prescribed neutron emission ratios \mymathtv{\xi_{\myCf}/\xi_{\myPuBe}\in\{\num{1},\num{1d1},\num{1d2}}\}. To ensure robust resampling across the extended dynamic range of event counts, the multinomial sampling probabilities \mymatht{\mathbf{p}} were defined proportional to the high-fidelity spectral templates \mymatht{\myvect{\uppsi}} (see \cref{sec:BayesTheory,sec:apptemplate}). For each configuration, we utilized \mymathtv{N_\text{rep}=\num{1d3}} realizations per grid point, with total event counts \mymatht{N^{(b)}_{\mathrm{event}}} distributed across a logarithmically spaced grid of \num{1d2} instances from \mymathtv{N^{(1)}_{\mathrm{event}}=1} to a maximum of \mymathtv{N^{(B)}_{\mathrm{event}} = (1+\xi_{\myCf}/\xi_{\myPuBe})\num{1d8}}. Bayesian evidence values for these realizations were computed via \cref{eq:logZb}, where the reference log-evidence \mymatht{\log\mathcal{Z}} and log-likelihood \mymatht{\log \mathcal{L}(\myvect{\uptheta};\mathcal{D},\mathcal{M})} were defined relative to the high-statistics dataset \mymathtv{\mathcal{D}\coloneqq\mathcal{D}^{(B)}} and retrieved using the nested-sampling procedure detailed in \cref{subsec:appevidence}. The statistical uncertainty in the posterior model probability \mymatht{p(\mathcal{M}_\mathrm{true}\mid\mathcal{D}^{(b)})} was then again quantified via Monte Carlo error propagation over the available realizations.

\subsection{\label{subsec:appig}Information Gain}

Building on earlier information theory work in Bayesian parameter inference \cite{Lindley1956,Luttrell1985,MacKay1992}, we continued our event-scale analysis by assessing the information gained about the true source model \mymatht{\mathcal{M}_\mathrm{true}} from the dataset \mymatht{\mathcal{D}^{(b)}}. For each resampled dataset \mymatht{\mathcal{D}^{(b)}}, we quantified the information gain as the Kullback–Leibler divergence (relative entropy) between the posterior and prior probabilities for the true model \cite{Kullback1951}:

\begin{equation}
\mathcal{IG}(\mathcal{M}_\mathrm{true}; \mathcal{D}^{(b)})
= p(\mathcal{M}_\mathrm{true}\vert\mathcal{D}^{(b)}) 
  \log_2\frac{p(\mathcal{M}_\mathrm{true}\vert\mathcal{D}^{(b)})}
             {p(\mathcal{M}_\mathrm{true})},
\label{eq:kld}
\end{equation}

\noindent where \mymatht{p(\mathcal{M}_\mathrm{true} \mid \mathcal{D}^{(b)})} is the posterior model probability obtained through \cref{eq:logpmodel}. This formulation quantifies the information gained about the true model from the observed data relative to the prior expectation in units of bits. The statistical uncertainty in \mymatht{\mathcal{IG}} was again estimated by propagating the Monte Carlo variability across the \mymatht{N_\mathrm{rep} = \num{1d3}} realizations of the resampled datasets. Averaging over these realizations yielded the expected information gain and its standard error, providing a robust measure of how strongly each dataset supports the true model under the competing hypothesis set \mymatht{\{\mathcal{M}_k\}_{k=1}^{K}}.

\section{\label{sec:appspec}Neutron Spectrometer}

The neutron spectrometer employed in this work is an organic glass scintillator (OGS) based system that has been characterized extensively in prior studies \cite{Lopez2022,Clarke2024a,Lopez2025}. OGS material offers a combination of high light output, fast scintillation response, and excellent pulse shape discrimination (PSD), making it particularly well suited for neutron spectroscopy across both recoil and TOF modalities \cite{Nguyen2021,Shin2019,Warburton2021}. The OGS composition used in this work consists of a 90:10 molecular blend of bis(9,9\allowbreak-di\-me\-thyl\allowbreak-9H\allowbreak-fluo\-ren\allowbreak-2\allowbreak-yl)\allowbreak{di\-phe\-nyl\-si\-lane}, \ce{C42H36Si}, and phe\-nyl\-tris(9,9\allowbreak-di\-me\-thyl\allowbreak-9H\allowbreak-fluo\-rene\allowbreak-2\allowbreak-yl)si\-lane, \ce{C51H44Si}, combined with a 0.2 wt\% 1,4\allowbreak-bis(2\allowbreak-methylstyryl)benzene, \ce{C24H22}, wavelength shifter (bis-MSB). Detailed information about the chemistry and synthesis of these compounds can be found in \myonlinecite{Carlson2016}.

The spectrometer consists of twelve OGS bars, each with dimensions of \mymatht{\num{6}\times\num{6}\times\qty{50}{\cubic\mm}} (see \cref{fig:setup}). All bars are wrapped in poly\-tetra\-fluo\-roe\-thyl\-ene (Teflon) tape to enhance diffuse optical reflection and improve light-collection uniformity. The OGS bars are coupled between two 64-pixel ArrayJ-60035-64P-PCB Onsemi silicon photomultiplier (SiPM) boards operated at \qty{29.6}{\V}. This dual-ended readout configuration allows for event reconstruction on the vertical axis of the OGS bars \cite{Ruch2016a,Lopez2022}. Prior work has demonstrated a time resolution of \qty{270}{\ps} at \myqtyrshort{200}{341}{\keVee} between two OGS bars \cite{Giha2021}. The spectrometer also features eight encapsulated cylindrical CeBr$_3$ crystals (\qty{6}{\mm} diameter and height) for gamma-ray spectroscopy and imaging, which are not relevant for this work. All scintillator–SiPM interfaces are coupled using \qty{0.5}{\mm} thick silicone rubber EJ-560 pads to ensure stable optical transmission. The system uses custom-made printed circuit boards (PCBs) to apply the voltage bias and readout the signals from the SiPM arrays \cite{Giha2017}. The signals from all channels are digitized using synchronized CAEN v1730 waveform digitizers (14-bit resolution, \qty{500}{\MHz} sampling rate , \qty{2}{\V} dynamic range). The CAEN v1730 modules were synchronized using a shared reference clock to maintain consistent timing across boards, enabling accurate reconstruction of inter-bar time differences for TOF analysis. The full detector assembly is enclosed in an aluminum dark box to suppress ambient light. A schematic photograph of the instrument and its mechanical layout is shown in \cref{fig:setup}.

\section{\label{sec:apppost}Data Reduction}

In this work, we adopt a benchmarked data reduction pipeline that has already been discussed in detail in our previous work \cite{Lopez2022,steinberger2020c}. In this subsection, we limit the discussion to key steps used for recoil and TOF spectroscopy. Further information on waveform processing, filtering, and uncertainty quantification is provided in \myonlinecitepl{Lopez2022,steinberger2020c}. 

Data are collected in full-waveform acquisition mode with a record length of \qty{800}{\ns}. Energy calibration was performed using {\myCsiso} measurements at the start and end of each experiment, following a calibration protocol described in \myonlinecite{Steinberger2021}. For each detected event, the signals recorded at each end of the organic glass scintillator bar are summed together if triggered within a \qty{20}{\ns} coincidence window. The start time of each detected event is determined by the average of the start times from the paired signals. Neutron and gamma-ray signals were separated using charge-integration PSD with an auto-slicing algorithm \cite{Brooks1960,Polack2015}. This PSD step is applied to all events and is common to both recoil and TOF spectroscopy.

\subsection{Recoil Spectroscopy} 
In recoil spectroscopy mode, we process all PSD-classified neutron events from individual OGS bars. For each event, the scintillation pulse produced by the neutron-induced proton recoil in OGS is integrated to obtain the pulse integral, which quantifies the light output. The calibrated light output is then expressed in electron-equivalent deposited energy to generate the recoil energy spectra.

\subsection{TOF Spectroscopy} 
TOF spectroscopy extends the analysis to multi-bar coincidence events to reconstruct the incident neutron energy. The procedure begins with the same PSD-classified neutron population used for recoil spectroscopy. From this set, candidate TOF events are selected by requiring a two-bar coincidence within a timing window of \myqtyrshort{0.137}{9}{\ns}, motivated by the geometrical constraints of the array and the selected neutron energy band. This filter isolates neutrons that undergo sequential inelastic scattering in two different bars while suppressing accidental coincidences. For each accepted coincidence event, the incident neutron energy \(E_0\) is then reconstructed as

\begin{equation}
E_0 = E_\text{dep,1} + E_\text{TOF}
\end{equation}

\noindent where \(E_\text{dep,1}\) denotes the energy deposited in the first scatter. This quantity is derived from the pulse integral light output and corrected for the non-proportional scintillation response of OGS using the Birks model \cite{Birks1951a,Lopez2025}. The second term, \(E_\text{TOF}\), represents the neutron energy inferred from the measured inter-bar flight time and the reconstructed interaction positions along each bar \cite{Lopez2022,steinberger2020c}. To mitigate systematic biases from gamma-ray leakage, pileup, and geometrically inconsistent scatter sequences, we impose an additional kinematic constraint

\begin{equation}
E_\text{dep,2} \le E_\text{TOF}
\end{equation}

\noindent where \(E_\text{dep,2}\) is the energy deposited in the second scatter. This requirement enforces that the second interaction cannot deposit more energy than is kinematically available from the neutron after the first interaction. Finally, we exclude events for which the relative uncertainty in the reconstructed neutron energy exceeds \qty{50}{\percent}, ensuring a high-quality dataset dominated by well-reconstructed trajectories. The reconstructed set of incident neutron energies $\{E_0\}$ is then used to generate the TOF neutron spectra presented in \cref{sec:MainResults}.

\section{\label{sec:apptemplate}Template Generation}

\begin{figure}[t]
\includegraphics[scale=1]{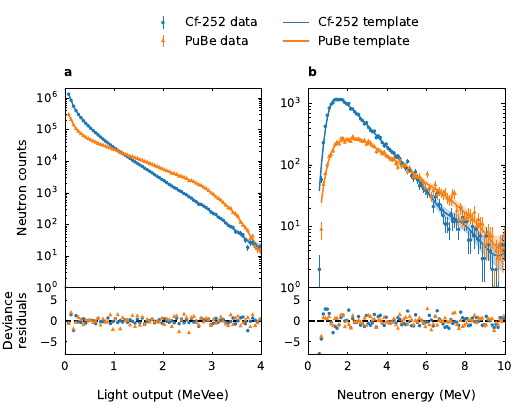}
\caption[Generalized additive model templates]{\label{fig:gamlearned}GAM‑based spectral templates for {\myCf} and {\myPuBe} together with the corresponding calibration data ({\mydbone} and {\mydbthree}, see \cref{tab:measurement}) for two spectroscopy modalities: (a) Recoil spectroscopy; (b) TOF spectroscopy. Measurement uncertainties are shown with coverage factor $k=1$, while template uncertainties are indicated as \qty{99}{\percent} prediction intervals. For better interpretability, the templates are scaled by the neutron emission rate of each source to represent absolute neutron counts.}
\end{figure}

As discussed in \cref{sec:BayesTheory}, we define the source model \mymatht{\mathcal{M}} in this work as the discrete ensemble of \mymatht{S\in\myNat} candidate neutron sources together with their associated template set \mymatht{\mathcal{T} = \{\myvect{\uppsi}_s\}^S_{s=1}}.  Each template \mymatht{\myvect{\uppsi}_s \in \myRealul{N}{+}} represents the normalized full-spectrum response of the spectrometric system with \mymatht{N\in\myNat} channels to the \mymatht{s}-th neutron source. Following a full-spectrum template matching approach, we then parameterize the forward mapping from any admissible parameter vector \mymatht{\myvect{\uptheta}\in\myRealu{M}} to the expected spectral response of a given spectrometric system \mymatht{\mathcal{M}(\myvect{\uptheta}) : \Uptheta \subseteq \mathbb{R}^M \mapsto \myNatu{N}} as the linear superposition of \mymatht{S\in\myNat} spectral templates \mymatht{\mathcal{M}(\myvect{\xi}) = \sum_{s=1}^{S} \xi_s \, \myvect{\uppsi}_s}, where \mymatht{\xi_s \in \mathbb{R}_{+}} denotes the corresponding neutron emission rate.

To generate the template bank required to evaluate the full power set of a {\myCf} and {\myPuBe} source considered in \cref{sec:MainResults}, we performed two additional single-source {\myCf} ({\mydbone}) and {\myPuBe} ({\mydbthree}) calibration experiments, as discussed in \cref{sec:Measurement}.  Adopting a data-driven strategy to minimize model bias, we generated the templates using a generalized additive model (GAM) implemented via the \mycode{pyGAM} package \cite{Serven2025}. For each source, we inferred a semi-parametric GAM from the calibration data using penalized iteratively reweighted least squares (\num{1d-5} tolerance). We employed a variable number of \mymatht{F} penalized B-spline basis functions ($F \in [30,70]$ for recoil and $F \in [2,12]$ for TOF spectroscopy), a Poisson log-link, and a regularization strength $\lambda_\text{GAM} \in [10^{-1},10^{3}]$. Both the number of basis functions $F$ and the regularization parameter $\lambda_\text{GAM}$ were selected through hyperparameter tuning. The learned spectral templates, together with the underlying calibration data, are shown in \cref{fig:gamlearned} for recoil and TOF spectroscopy.

As discussed in \cref{sec:Measurement}, the source position and source casing were deliberately varied between the template-generation and identification experiments to demonstrate the robustness and generalization capability of the derived templates. To avoid any significant bias in the retrieval of the source set between the templates and the experimental data, we restricted the energy bands used for identification to \qtyrange{0.1}{3.5}{\MeVee} for recoil spectroscopy (bin width \mymatht{\Delta E=\qty{0.05}{\MeVee}}, \mymatht{N=68} channels) and \qtyrange{2}{10}{\MeV} for TOF spectroscopy (bin width \mymatht{\Delta E=\qty{0.1}{\MeV}}, \mymatht{N=80} channels). A sensitivity study confirmed that the identification results are stable under \qty{10}{\keV} variations of these band limits (see Supplemental Material \cite{zotero-item-8041}).



%

\end{document}